\documentclass[aps, prd, 10pt, twocolumn, superscriptaddress,noshowpacs, preprintnumbers, longbibliography,nofootinbib,bibnotes,floatfix]{revtex4-2}
\pdfoutput=1 

\usepackage{amsmath}
\usepackage{amsfonts}
\usepackage{amssymb}
\usepackage{mathtools}
\usepackage{graphics}
\usepackage{tabularx}
\usepackage[export]{adjustbox}
\usepackage{multirow}
\usepackage{citesort}
\usepackage{graphicx}
\usepackage{url}
\usepackage{soul}
\usepackage{physics}
\usepackage{bm}
\usepackage[dvipsnames]{xcolor}
\usepackage[utf8]{inputenc}
\usepackage[normalem]{ulem}

\usepackage{tikz,xcolor}
\usepackage{hyperref}

\definecolor{lime}{HTML}{A6CE39}
\DeclareRobustCommand{\orcidicon}{\hspace{-1mm}
 \begin{tikzpicture}
 \draw[lime, fill=lime] (0,0) 
 circle [radius=0.16] 
 node[white] {{\fontfamily{qag}\selectfont \tiny \,ID}};
 \draw[white, fill=white] (-0.0525,0.095) 
 circle [radius=0.007];
 \end{tikzpicture}
 \hspace{-3mm}
}

\foreach \x in {A, ..., Z}{\expandafter\xdef\csname orcid\x\endcsname{\noexpand\href{https://orcid.org/\csname orcidauthor\x\endcsname}
 {\noexpand\orcidicon}}
}

\begin{document}

\title{Diagnosing electron-neutrino lepton number crossings in core-collapse supernovae:\texorpdfstring{\\ A comparison of methods}{}}

\author{Marie Cornelius\orcidA{}}
\affiliation{Niels Bohr International Academy and DARK, Niels Bohr Institute, University of Copenhagen, Blegdamsvej~17, 2100 Copenhagen, Denmark}
\author{Irene Tamborra\orcidB{}}
\affiliation{Niels Bohr International Academy and DARK, Niels Bohr Institute, University of Copenhagen, Blegdamsvej~17, 2100 Copenhagen, Denmark}
\author{Malte Heinlein\orcidC{}}
\affiliation{Max-Planck-Institut f\"{u}r Astrophysik, Karl-Schwarzschild-Str.~1, 85748 Garching, Germany}
\affiliation{Technische Universit\"{a}t M\"{u}nchen, TUM School of Natural Sciences, Physics Department, James-Franck-Str.~1, 85748 Garching, Germany}
\author{Hans-Thomas Janka\orcidD{}}
\affiliation{Max-Planck-Institut f\"{u}r Astrophysik, Karl-Schwarzschild-Str.~1, 85748 Garching, Germany}

\date{\today}

\begin{abstract}
Fast neutrino flavor conversion may impact the explosion mechanism and nucleosynthesis in core-collapse supernovae. A necessary condition for fast flavor conversion is the presence of crossings in the angular distribution of the electron-neutrino lepton number (ELN) crossing. Because of the computational costs, flavor-dependent angular distributions are not computed by the vast majority of state-of-the-art hydrodynamical simulations; instead, angular distributions are reconstructed employing approximate methods in post-processing.
In this work, we evaluate the performance of four methods adopted to diagnose the existence of ELN crossings. For selected post-bounce times, we extract the fluid and thermodynamic properties from spherically symmetric supernova simulations for an $18.6 M_\odot$ progenitor, testing cases with and without muons as well as with and without mixing-length treatment of proto-neutron star convection. We compare the occurrence of crossings in the angular distributions obtained by solving the Boltzmann equations with those in distributions reconstructed from angular moments of our Boltzmann solutions by using the maximum entropy and Minerbo schemes, and also with crossings identified via a polynomial weighting function applied to the angular moments.
Our results show that the polynomial method and the Minerbo closure scheme have severe limitations. The maximum entropy approach captures most of the forward crossings, although it fails to reproduce or misidentifies crossings in a subset of our models. 
These findings highlight the need for robust modeling of the neutrino angular properties in order to assess the impact of flavor conversion on the supernova mechanism.
\end{abstract}

\maketitle

\section{Introduction}

Core-collapse supernovae (SNe) mark the explosive death of massive stars, with neutrinos playing a central role in the SN mechanism, driving the transfer of energy and lepton number--see, e.g., Refs.~\cite{Janka:2025tvf,Burrows:2020qrp,Tamborra:2024fcd,Vitagliano:2019yzm} for recent reviews on the topic. According to the delayed neutrino-heating mechanism~\cite{Bethe:1985sox}, neutrinos emitted from the hot proto-neutron star (PNS) deposit energy behind the stalled shock wave, reviving the latter and enabling the SN explosion. Neutrinos also drive the nucleosynthesis of heavy elements, with the relative fluxes of electron neutrinos and antineutrinos contributing to determine the neutron-to-proton ratio~\cite{1996ApJ...471..331Q,Hoffman:1996aj}; for a review, see Ref.~\cite{Fischer:2023ebq}. 
Hydrodynamical SN simulations assume that neutrino flavor does not change while neutrinos stream through the SN medium~\cite{Mezzacappa:2020oyq}. This approximation may be justified in the free-streaming regime, where the feedback of neutrino physics on the stellar medium may be negligible. However, such simplification does not hold during neutrino decoupling when the flavor physics can have a non-negligible impact on the supernova medium. In particular, flavor conversion is affected by coherent forward scattering on other neutrinos or on matter~\cite{Duan:2010bg,Mirizzi:2015eza,Tamborra:2020cul,Johns:2025mlm,Volpe:2023met}. This phenomenon can alter the SN explosion mechanism as well as the observable multi-messenger signals~\cite{Ehring:2023lcd,Ehring:2023abs,Nagakura:2023mhr,Wang:2025nii}.
The phenomenology induced by neutrino self-interactions is very rich and remains poorly understood~\cite{Duan:2010bg,Mirizzi:2015eza,Tamborra:2020cul,Johns:2025mlm,Volpe:2023met}. Fast flavor conversion can take place in the limit of vanishing vacuum frequency, with a characteristic scale depending on the local neutrino number density. Fast flavor conversion can occur if a crossing exists in the angular distribution of the electron lepton number of neutrinos (ELN)~\cite{Izaguirre:2016gsx,Morinaga:2021vmc,Padilla-Gay:2021haz,2023PhRvD.107l3024F}.

In order to improve our understanding of the spatial regions in the SN core and the phases of the SN evolution mostly affected by fast flavor conversion, a large body of literature focuses on searching for ELN crossings in hydrodynamical SN simulations.
However, except for a few cases, e.g.~Refs.~\cite{Tamborra:2017ubu,Sumiyoshi:2012za,Akaho:2020xgb,Nagakura:2017mnp,Fischer:2009af}, hydrodynamical SN simulations do not compute the neutrino angular distributions, due to the high computational costs. Most neutrino-hydrodynamic simulations of SNe instead evolve the lowest angular moments of the neutrino distribution functions--for example, the neutrino field is described using the zeroth (number density) and the first (flux) moments--and closure schemes are employed to deal with the approximations introduced by neglecting higher angular moments; see, e.g.~Ref.~\cite{Mezzacappa:2020oyq} for a recent review on neutrino transport techniques. 
One of the most popular closures is the maximum entropy method, which assumes that the neutrino angular distribution is such to maximize entropy~\cite{MINERBO1978541,Cernohorsky:1994yg}. While closure prescriptions offer a good solution for simplifying energy-momentum transport, they are not adequate for tracking the evolution of the neutrino field in momentum space. Alternatively, based on the first few angular moments, a weighting function can be constructed to highlight the crossing region~\cite{Abbar:2020fcl} or one could rely on the instability of the zeroth Fourier mode~\cite{Dasgupta:2018ulw}.

The first search for ELN crossings based on a suite of spherically symmetric SN models--with angular distributions computed consistently in the hydrodynamical simulations--led to negative results~\cite{Tamborra:2017ubu}.
However, follow-up work based on multi-dimensional SN models found that ELN crossings should exist in both forward and backward directions of neutrino propagation in the convective layer of the PNS~\cite{DelfanAzari:2019tez,Akaho:2022zdz,Akaho:2023brj,Glas:2019ijo}.
In the trapped regime, the angular distributions of all flavors are isotropic. ELN crossings arise due to rapid deleptonization, which favors a shift of beta-equilibrium toward negative chemical potentials for electron neutrinos in the regions where electron neutrinos and antineutrinos are in chemical equilibrium with the stellar medium. This effect efficiently occurs in the PNS convective layer, therefore the regions of ELN crossings are affected by convective flow asymmetries in multi-dimensional SN simulations.
The angular crossings tend to be narrow, reflecting the local effects of convection-driven asymmetries. At larger radii, closer to the neutrinospheres and outside of those, the angular distributions of electron antineutrinos are more forward peaked than those of electron neutrinos because the former decouple at radii smaller than the latter; this can lead to ELN crossings in the decoupling region~\cite{Nagakura:2019sig,Nagakura:2021hyb,Akaho:2022zdz,Akaho:2023brj}. 
ELN crossings have also been found in the pre-shock region, before the revival of the stalled shock~\cite{Nagakura:2019sig,Morinaga:2019wsv,Harada:2021ata,Akaho:2023brj,Abbar:2020qpi,Capozzi:2020syn}.
In fact, electron antineutrinos, having a higher average energy than electron neutrinos, experience stronger backward scattering off nuclei. This results in more electron antineutrinos being deflected into the backward direction than electron neutrinos, inducing backward crossings in the ELN distribution. However, the backward neutrino flux in the pre-shock region is negligible after neutrinos decouple from the matter background; thus the presence of backward ELN crossings might be negligible for what concerns flavor evolution.

In this paper, we adopt a spherically symmetric core-collapse SN simulation~\cite{Garching_CCSN_archive}, featuring a progenitor mass of $18.6\ M_{\odot}$ and a Lattimer and Swesty nuclear equation of state with an incompressibility modulus equal to $220$~MeV 
(LS220)~\cite{Lattimer:1991nc}. For comparison, we also consider SN models with identical properties, but including muons and a mixing-length treatment for PNS convection. For selected post-bounce times, we focus on the spatial region surrounding neutrino decoupling and compute the angular distributions of all neutrino flavors solving the Boltzmann equations, following the approach presented in Refs.~\cite{Shalgar:2023aca,Shalgar:2024gjt,Cornelius:2025nvd}.
We then search for ELN crossings and compare our findings to the ones obtained by employing approximate schemes based on the first few moments derived from the neutrino distributions computed by our Boltzmann solutions. 

We build on Ref.~\cite{Johns:2021taz}, which pointed out that the limited information provided by the first few moments may hamper the search for angular crossings. Our goal is to assess to what extent the regions where flavor conversion may take place can be identified using moment-based schemes by means of a systematic comparison with the Boltzmann solution.

Our work is organized as follows. The SN models employed in this paper are outlined in Sec.~\ref{Subsec:models}. Section~\ref{Sec:methods} introduces the methods adopted to compute the neutrino angular distributions, i.e.~the solution of the Boltzmann equations and the approximate moment-based approaches such as the maximum entropy and the Minerbo closure schemes, as well as the alternative polynomial method.
In Sec.~\ref{Sec:results}, we present a comparison of the performance of these methods to identify ELN crossings across six post-bounce times for our benchmark SN models along with a discussion of our findings. Conclusions are presented in Sec.~\ref{Sec:conclusions}. In addition, Appendix~\ref{Sec:appendix} provides additional details on the functions adopted in the polynomial method. 

\section{Supernova models}
\label{Subsec:models}
Our reference case is a spherically symmetric core-collapse SN simulation~\cite{Garching_CCSN_archive} with a mass of $18.6 M_{\odot}$, LS220 nuclear equation of state~\cite{Lattimer:1991nc}, and without muons as well as PNS convection; we refer the interested reader to Ref.~\cite{Fiorillo:2023frv} for additional details. For simplicity, hereafter, we name this Model 1 (without muons and without convection). In order to assess the role of muons in the stellar medium and PNS convection on the formation of ELN crossings, we also consider two additional SN models employing a mixing-length treatment for PNS convection~\cite{2012PhRvL.108f1103R,Mirizzi:2015eza}: Model 2 (without muons and with convection) and Model 3 (with muons and with convection). The baryonic mass of the neutron star is $1.61 M_{\odot}$ for Models 1 and 2, and $1.62 M_{\odot}$ for Model 3. In what follows, unless otherwise specified, we consider two neutrino flavors: electron neutrinos $\nu_e$ and heavy-lepton neutrinos $\nu_x$ with $x=\mu$ or $\tau$, and the corresponding antineutrinos. Note that the SN models including muons take into account muon effects in the equation of state of the stellar plasma as well as in the neutrino transport via neutrino-muon interactions. In contrast, muon reactions are not included in the Boltzmann neutrino transport applied in the present work, as discussed in Sec.~\ref{Subsec:Boltzmann}.
For each SN model, we consider six post-bounce time snapshots: $0.05$, $0.1$, $0.5$, $0.75$, $1$, and $3$~s, covering both the PNS accretion and cooling phases. An overview of the SN models introduced above is provided in Table~\ref{Tab:SN_models}.

\begin{table}[]
\renewcommand{\arraystretch}{1.2}
\setlength{\tabcolsep}{5pt}
\caption{Characteristic properties and radial ranges adopted to solve the Boltzmann equations for SN Models 1, 2 and 3.}
\begin{tabular}{|l|cc|ccc|}
\hline
 & \multicolumn{2}{c|}{Model 1} & \multicolumn{2}{c|}{Model 2} & Model 3 \\ \hline \hline
Muons & \multicolumn{2}{c|}{No} & \multicolumn{2}{c|}{No} & Yes \\ \hline
PNS convection & \multicolumn{2}{c|}{No} & \multicolumn{2}{c|}{Yes} & Yes \\ \hline \hline
$t_{\mathrm{pb}}$ & $r_{\min}$ & $r_{\max}$ & \multicolumn{2}{c}{$r_{\min}$} & $r_{\max}$ \\ \hline
$0.05$ s & $25$ km & $175$ km & \multicolumn{2}{c}{$25$ km} & $175$ km \\
$0.10$ s & $25$ km & $135$ km & \multicolumn{2}{c}{$25$ km} & $135$ km \\
$0.50$ s & $18$ km & $37$ km & \multicolumn{2}{c}{$15$ km} & $37$ km \\
$0.75$ s & $16$ km & $31$ km & \multicolumn{2}{c}{$13$ km} & $31$ km \\
$1.00$ s & $14$ km & $29$ km & \multicolumn{2}{c}{$12$ km} & $29$ km \\
$3.00$ s & $12$ km & $21$ km & \multicolumn{2}{c}{$10$ km} & $21$ km \\ \hline
\end{tabular}
\label{Tab:SN_models}
\end{table}

The neutrino transport module of the {\tt PROMETHEUS-VERTEX} neutrino-hydrodynamics code employed in our benchmark simulations integrates the velocity-dependent neutrino energy and momentum equations. These angular-moment equations are then closed by variable Eddington factors deduced from corresponding solutions of Boltzmann transport equations that were iterated for convergence and consistency with the moments obtained from the angular-moment equations~\cite{Rampp:2002bq}. Since the computed information of the neutrino angular distributions is not available for the considered suite of SN models, for each of the selected post-bounce times, we compute the angular distributions of neutrinos as solutions of the Boltzmann equations for static SN backgrounds (i.e.~$v = 0$, with $v$ being the velocity of the stellar medium), reconstructing and approximating them through the different tested moment-based methods.

\section{Methods to construct the angular distribution of the electron neutrino lepton number}
\label{Sec:methods}
In this section, we introduce the Boltzmann equations as well as the moment-based methods adopted to search for ELN crossings.

\subsection{Boltzmann neutrino transport} 
\label{Subsec:Boltzmann}
The angular distributions of neutrinos and antineutrinos can be computed by solving the Boltzmann equations. This method does not truncate the infinite set of angular-moment equations of the Boltzmann equations on the level of the first few moment equations (cf.~Sec.~\ref{Subsec:moment_methods}) and therefore, in principle, preserves the angular information. 

For fixed post-bounce time, we follow the approach outlined in Refs.~\cite{Shalgar:2023aca,Shalgar:2024gjt} and extract the fluid properties such as the baryon density, chemical potentials, and electron fraction from the SN simulations in Table~\ref{Tab:SN_models} as functions of radius $r$ and time $t$. The Boltzmann equations for neutrinos and antineutrinos also depend on the propagation angle (relative to the radial direction, $\mu = \cos\theta$) and neutrino energy $E$, respectively: 
\begin{eqnarray}
\label{Eq:KE}
 \left(\pdv{t}+\vec c \cdot \vec \nabla\right) \rho &=& \mathcal{C}[\rho, \bar\rho]\ ,\\
    \left(\pdv{t}+\vec c \cdot \vec \nabla\right) \bar\rho &=& \mathcal{C}[\rho, \bar\rho]\ ,
\label{Eq:KE1}
\end{eqnarray}
where $\vec c$ is the speed of light for the neutrino propagation and $\rho = \rho(r,\mu,E,t)$ ($\bar\rho = \bar\rho(r,\mu,E,t)$) is the (anti)neutrino density matrix, whose diagonal terms represent the occupation numbers of (anti)neutrinos of different flavors. The energy- and angle-integrated occupation numbers, $\int \rho_{ii}\mathrm{d}E\mathrm{d}\mu$, represent the local $\nu_i$ number densities [$n_{\nu_i}(r)$]. The gradient term on the left-hand side of the equations describes (anti)neutrino propagation, while the operator $\mathcal{C}$ accounts for (anti)neutrino interactions with matter (i.e., beta processes, pairwise annihilation and production, bremsstrahlung, and direction-changing interactions). The collision term is modeled following Ref.~\cite{OConnor:2014sgn}, as detailed in Appendix A of Ref.~\cite{Shalgar:2023aca} and Sec.~2 of Ref.~\cite{Shalgar:2024gjt}.

For each post-bounce time, we solve the Boltzmann equations within a spherical shell that spans from $r_{\min}$ to $r_{\max}$, encompassing the neutrino decoupling region, until a steady-state solution is achieved~\cite{Cornelius:2025nvd,Shalgar:2023aca,Shalgar:2022rjj}. 
The selected radial range for each post-bounce time is provided in Table~\ref{Tab:SN_models}. The minimum radius is chosen such that the decoupling radius for the smallest neutrino energy is included in the simulation.
At $r_{\min}$, the angular distributions of (anti)neutrinos are isotropic, while neutrinos effectively free-stream at $r_{\max}$, with their angular distributions having a negligible backward flux. The energy range spans from $1$ to $100$~MeV. Equations~\ref{Eq:KE} and \ref{Eq:KE1} are solved numerically on a grid of $150$ radial bins, $300$ $\mu$ angular bins, and $100$ energy bins, all linearly spaced. 

The ELN angular distribution using the density matrix formalism is
\begin{equation}
  G(r,\mu) = \int \mathrm{d}E \left[\rho_{ee} - \bar\rho_{ee}\right] \ ,
\end{equation}
where $\rho_{ee}=\rho_{ee}(r,\mu,E)$ and $\bar\rho_{ee}=\bar\rho_{ee}(r,\mu,E)$ represent the $\nu_e$ and $\bar\nu_e$ occupation numbers, respectively. 
An ELN crossing occurs at a given radius for $G(r,\mu) = 0$.

\begin{figure*}
  \centering
  \includegraphics[width=0.99\linewidth]{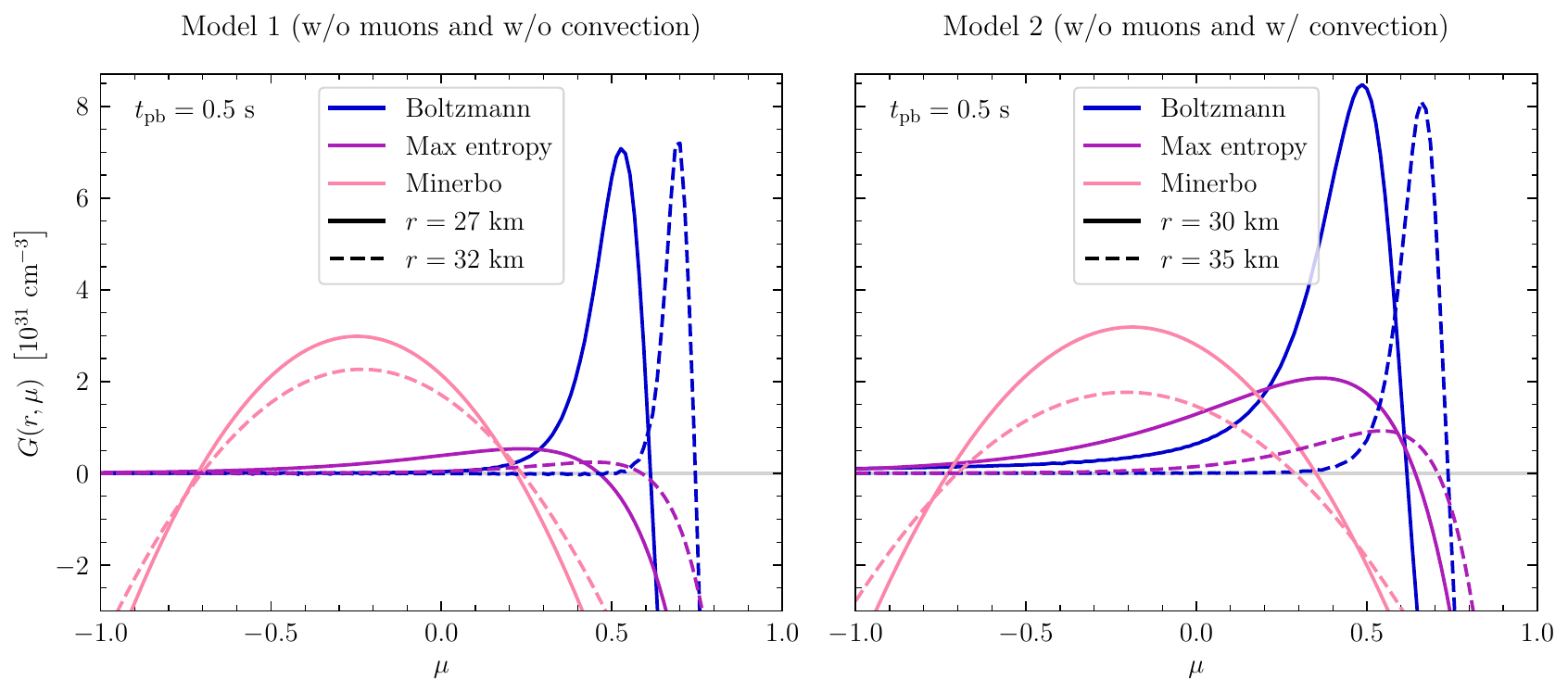}
  \caption{ELN distributions obtained solving the Boltzmann equations (blue), using the maximum entropy distributions (purple), and the Minerbo closure (pink). The gray horizontal line marks $G(r,\mu)=0$ to guide the eye. \textit{Left panel}: ELN distributions for Model 1 (w/o muons and convection) extracted at $27$~km (solid) and $32$~km (dashed). The energy-integrated $\nu_e$ decoupling radius is located at $22.5$~km. \textit{Right panel}: ELN distributions for Model 2 (w/o muons and w/ convection) extracted at $30$~km (solid) and $35$~km (dashed). The energy-integrated decoupling radius for $\nu_e$'s is located at $25.5$~km.
  }
  \label{fig:ang_dists}
\end{figure*}
The left (right) panel of Fig.~\ref{fig:ang_dists} displays examples of $G(\mu)$ computed at two different radii for Model 1 (Model 2). In both cases, we focus on $t_{\mathrm{pb}}=0.5$~s. As neutrinos travel outward, their distributions become increasingly forward peaked. Notably, electron antineutrinos decouple at smaller radii than electron neutrinos do since they have weaker interaction rates with matter in the SN core. This leads to the development of an ELN crossing for $\mu > 0$.

Physically motivated ELN crossings in the backward direction may not lead to relevant flavor conversion since the tails of the angular distributions of neutrinos of different flavors are approximately identical for $\mu \lesssim 0$. Hence, we choose to focus on ELN crossings in the forward direction, as these appear when the ELN angular distribution has a non-trivial shape and are of relevance for flavor conversion.
Because of the finite angular grid size in the Boltzmann solver, small numerical fluctuations can arise in the angular distributions. This may cause oscillations around $G(\mu)=0$, resulting in multiple spurious crossings. This issue is especially pronounced when the neutrino flux is negligible, such as in the backward direction of the distributions in Fig.~\ref{fig:ang_dists}. Here, numerical fluctuations can cause the angular distribution to become negative for one or both of the flavors, which is unphysical. Even when we restrict our search to forward crossings, the angular distributions at large radii may exhibit small fluxes for $0 < \mu < 0.5$, which might lead to unphysical crossings.
In order to avoid the misinterpretation of spurious ELN crossings caused by numerical artifacts, we apply the following criteria:
\begin{itemize}
     \item The electron neutrino and antineutrino angular distributions are positive at the crossing and for the angular bins immediately before and after the crossing.
     \item The angular bin immediately after the bin closest to $G(\mu)=0$ has opposite sign compared to the bin just before the crossing.
     \item The two consecutive angular bins after the bin closest to $G(\mu)=0$ have the same sign, and the two bins before $G(\mu)=0$ also share the same sign.
     \item To ensure that the angular distributions of electron (anti)neutrinos are forward peaked--with the electron antineutrino distribution being more forward-peaked than that of electron neutrinos, consistent with their earlier decoupling--we request that $\int d\mu G(\mu<0)$ should not be greater than the integral of $G(\mu)$ from $\mu=0$ to the angular index of the crossing.
     \item In order to avoid that any unphysical crossings appear near the upper boundary because of the specific choice of $r_{\rm max}$, we do not consider the last two radial bins in the search for ELN crossings.
\end{itemize}

\subsection{Moment-based methods}
\label{Subsec:moment_methods}
In lieu of computing the neutrino distributions with high angular resolution in momentum space through the solution of the Boltzmann equations, one can utilize approximate methods that rely on the first two angular moments. These are the energy-integrated neutrino number density and number flux and are given by
\begin{eqnarray}
  n_{\nu_i}(r) &=& \frac{2\pi}{(hc)^3}\int \mathrm{d}E E^2 \int \mathrm{d}\mu f_{\nu_i}(r,\mu,E)\ ,\\
  F_{\nu_i}(r) &=& \frac{2\pi c}{(hc)^3}\int \mathrm{d}E E^2 \int \mathrm{d}\mu~\mu f_{\nu_i}(r,\mu,E) \ ,
\end{eqnarray}
with $f_{\nu_i}(r,\mu,E)$ being the occupation number (or phase-space density) for the neutrino species $\nu_i$, which can be extended to the density matrix $\rho$ introduced in Sec.~\ref{Subsec:Boltzmann}, if one wants to account for flavor coherence in the oﬀ-diagonal elements.
Here, we consider $i=e$ or $x$ (i.e., without distinguishing between $\mu$ and $\tau$ neutrinos) in the solution of our Boltzmann equations. 

The M1 transport scheme employs a closure relation to approximate higher-order moments and reconstruct the angular distribution of neutrinos. Such a closure could be the analytical Minerbo closure~\cite{MINERBO1978541}.
Another widely used method relies on the assumption of maximum entropy~\cite{MINERBO1978541,Cernohorsky:1994yg}, which can be used to reconstruct the neutrino angular distributions and as a closure scheme. Alternatively, the polynomial method leverages on the first moments to highlight the presence of a crossing in the ELN angular distribution~\cite{Abbar:2020fcl}. In the following, we illustrate each of these approaches. 

\subsubsection{Maximum entropy method}
\label{Subsubsec:max_entropy}
The maximum entropy method is a widely used closure scheme adopted in neutrino transport, according to which the sequence of higher-order angular moments is cut by expressing the relevant ones entering the moment equations by a finite set of lower-order moments (just the first two for M1 transport)~\cite{MINERBO1978541,Cernohorsky:1994yg}.
The energy-integrated neutrino angular distributions can be reconstructed as follows~\cite{Richers:2022dqa,Cernohorsky:1994yg}: 
\begin{equation}
  f^{\mathrm{ME}}_{\nu_i}(r,\mu) = \frac{n_{\nu_i}(r)}{4\pi}\frac{Z}{\sinh(Z)}e^{Z\mu} \ ,
\end{equation}
where $Z$ is determined by solving the Langevin equation
\begin{equation}
  \tilde{f}_{\nu_i}(r) = \coth(Z)-\frac{1}{Z} \ ,
\end{equation}
and the dimensionless flux factor is defined as
\begin{equation}
  \tilde{f}_{\nu_i}(r)=\left|\frac{F_{\nu_i}(r)}{c n_{\nu_i}(r)}\right| \ .
  \label{Eq:flux_factor}
\end{equation}
We then compute the first two moments ($n_{\nu_i}$ and $F_{\nu_i}$) from the angular distributions obtained from our Boltzmann transport solver.
The ELN angular distribution is then given by
\begin{equation}
    G(r,\mu)=f^{\mathrm{ME}}_{\nu_e}(r,\mu) - f^{\mathrm{ME}}_{\bar\nu_e}(r,\mu) \ .
\end{equation}

The purple lines in Fig.~\ref{fig:ang_dists} represent the ELN angular distributions computed with the maximum entropy method for SN Model 1 (on the left) and Model 2 (on the right) at $t_{\mathrm{pb}}=0.5$ s.
Comparing the blue lines (Boltzmann solution) with the purple ones, the maximum entropy distributions are noticeably smoother, lacking the pronounced peaks of the Boltzmann distributions in the forward direction. 
In Models 1 and 2, the maximum entropy distributions have crossings for both radii. However, it is important to note that the shape of the reconstructed ELN distributions and the slope of the ELN crossings are different from the ones obtained through the Boltzmann equations; these differences have dramatic implications on flavor conversion physics~\cite{Padilla-Gay:2021haz,Tamborra:2020cul}.
This comparison highlights that angular distributions that are highly forward-peaked are challenging to reconstruct when only the first few moments are employed. 

\subsubsection{Minerbo closure method}
\label{Subsubsec:Minerbo}
In hydrodynamical simulations employing the M1 scheme, the zeroth and first moments of the neutrino distribution function--corresponding to the number density and flux--are solved directly, while the second moment ($M^2_{\nu_i}$) and third moment ($M^3_{\nu_i}$) of the number distribution are approximated using closures.
The analytical Minerbo closure is based on the maximum entropy principle~\cite{Just:2015fda,MINERBO1978541}. It relates the second moment to the flux factor by means of the following relation:
\begin{equation}
  \chi_{\mathrm{Minerbo}}(\tilde{f})=\frac{1}{3}+\frac{1}{15}\left(6\tilde{f}^2-2\tilde{f}^3+6\tilde{f}^4\right)\ ,
\label{Eq:chi_closure}
\end{equation}
where $M^2_{\nu_i}(r)\equiv \chi_{\mathrm{Minerbo}}(\tilde{f}) n_{\nu_i}(r)$ and $\tilde{f} \equiv \tilde{f}_{\nu_i}(r)$, which are computed from the first two angular moments derived from the solutions of the Boltzmann equations.
The angular distribution is then reconstructed by assuming a quadratic distribution of the form $f^{\mathrm{Minerbo}}_{\nu_i}(\mu)=a_0+a_1\mu+a_2\mu^2$. The coefficients $a_0,a_1,a_2$ are determined using the moments. 

Figure~\ref{fig:ang_dists} shows in pink the ELN distributions obtained by adopting the Minerbo closure. For Models 1 and 2, both forward and backward crossings exist.
Compared to the Boltzmann solution shown in blue, the Minerbo distributions exhibit peaks shifted toward negative $\mu$ instead of $\mu \gtrsim 0.5$. 
It is important to note that the Minerbo angular distributions are unphysical in some angular regions, exhibiting negative number densities for $\mu \in [-0.7, 0.1]$ at $27$~km and $\mu \in [-0.7, 0.2]$ at $32$~km for Model 1, and for $\mu \in [-0.7, 0.0]$ at $30$~km and $\mu \in [-0.7, 0.1]$ at $35$~km for Model 2. These artifacts suggest that the Minerbo closure is not suitable for identifying crossings, nor inferring the ELN distributions.

\subsubsection{Polynomial method}
\label{Subsubsec:polynomial}
According to the polynomial method~\cite{Abbar:2020fcl}, the angular distributions feature an ELN crossing, if there exists a function $\mathcal{F}(\mu)>0$ for $\mu \in [-1,1]$ such that
\begin{equation}
  I_{\mathcal{F}}(r)I_0(r) < 0 \ ,
\end{equation}
where
\begin{equation}
  I_{\mathcal{F}}(r) = \int_{-1}^{1}d\mu \mathcal{F}(\mu)G(r,\mu) \ ;
\end{equation}
the ELN angular distribution is 
\begin{equation}
  G(r,\mu) = \frac{2\pi}{(hc)^3}\int \mathrm{d}E E^2 \left[f_{\nu_e}(r,\mu,E) - f_{\bar\nu_e}(r,\mu,E)\right] \ .
\end{equation}
The polynomial function has therefore the role of highlighting the region of the ELN crossing to make the latter better detectable.

The angular moments are defined as
\begin{equation}
  I_n(r) = \int_{-1}^{1}d\mu \mu^n G(r,\mu) \ ,
\end{equation}
with $I_0(r) = n_{\nu_e}(r) - n_{\bar\nu_e}(r)$. The integral $I_{\mathcal{F}}(r)$ can be discretized into a sum of moments $I_n(r)$ such that $I_{\mathcal{F}}(r) = \sum_{n=0}^{N}a_nI_n(r)$
where $\mathcal{F}(\mu)$ is expanded as a polynomial
\begin{equation}
  \mathcal{F}(\mu)=\sum_{n=0}^{N}a_n\mu^{n} \ ;
\end{equation}
here, $N$ represents the highest order of moments available.
By selecting the coefficients $a_n$, $I_{\mathcal{F}}(r)$ acquires a sign opposite to that of $I_0(r)$, allowing one to identify the existence of an ELN crossing.

Following the procedure outlined in Ref.~\cite{Abbar:2020qpi}, the second and third moments can be calculated by means of Eq.~\ref{Eq:chi_closure} and the Minerbo closure relation of the third moment:
\begin{equation}
\begin{split}
  q_{\mathrm{Minerbo}}(\tilde{f})=\frac{\tilde{f}}{75}\Bigl( 45+10\tilde{f}-12\tilde{f}^2-12\tilde{f}^3
  \\+38\tilde{f}^4-12\tilde{f}^5+18\tilde{f}^6 \Bigr) \ ,
\end{split}
\end{equation}
with $M^3_{\nu_i}(r)\equiv q_{\mathrm{Minerbo}}(\tilde{f})cn_{\nu_i}(r)$. The fundamental neutrino moments $n_{\nu_i}$ and $F_{\nu_i}$ used in our application of the polynomial method are derived from our solutions of the Boltzmann equations.
Relying on the first four moments, we construct $\mathcal{F}(\mu)$ as a linear, quadratic, or cubic polynomial. The local minimum of each polynomial can be strategically placed near an expected peak of the ELN distribution (for $\mu>0$) to improve the likelihood of identifying forward crossings. 

The weighting function chosen for the polynomial method can be customized to enhance the detection of forward or backward crossings. In this work, we focus on forward crossings; therefore, we select the weighting functions
$\mathcal{F}(\mu)$ accordingly. The choice of polynomial functions used to identify crossings is detailed in Appendix~\ref{Sec:appendix}.
However, it is important to note that, since the angular distributions and the ELN crossing locations are not known a priori, the polynomial method does not inherently distinguish between forward and backward crossings, nor does it quantify the strength or width of a crossing.

\section{Search for electron lepton number crossings}
\label{Sec:results}

\begin{figure*}
  \centering
  \includegraphics[width=0.92\textwidth]{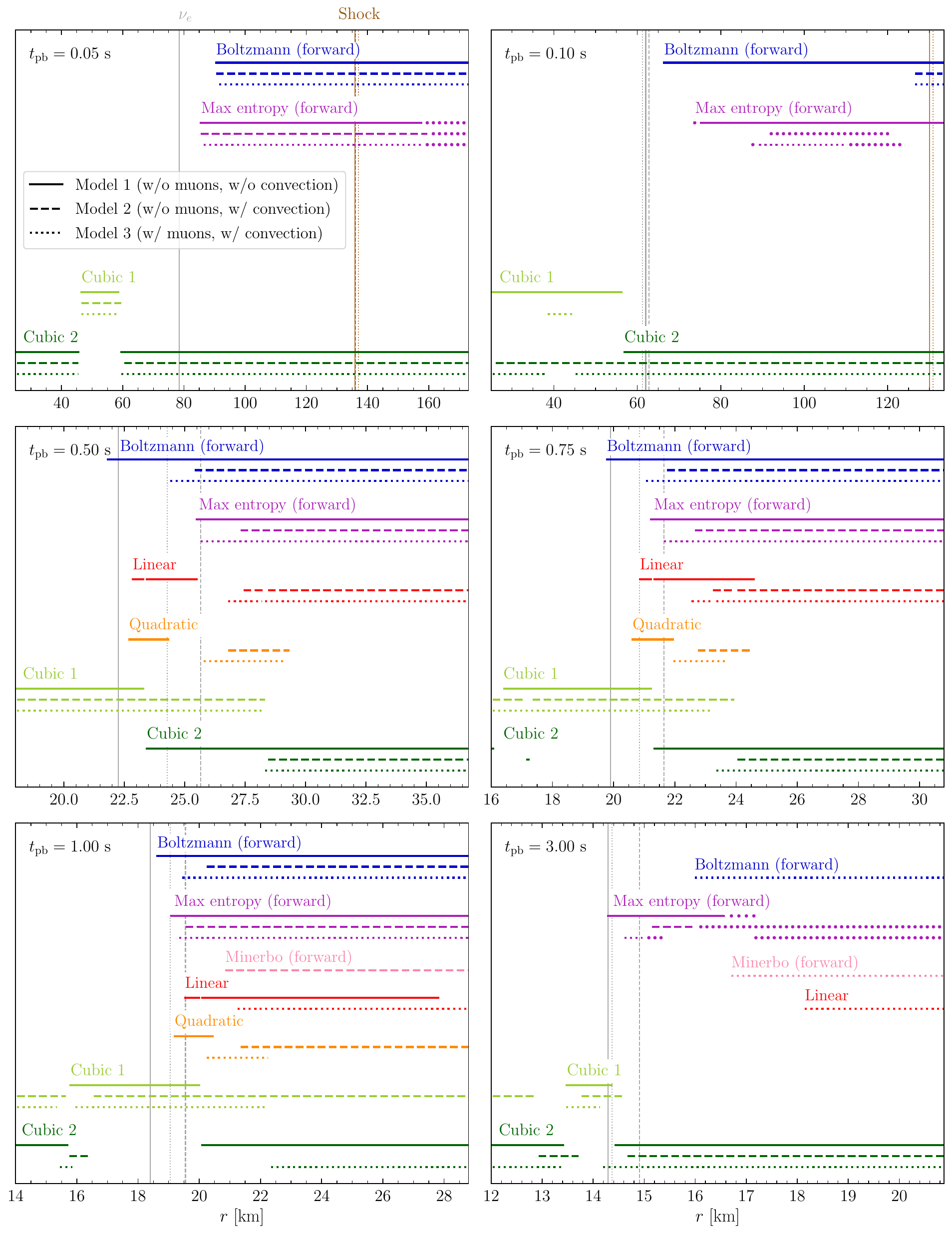}
  \caption{Locations of ELN crossings for the SN Model 1 (w/o muons and w/o convection, solid), Model 2 (w/o muons and w/ convection, dashed), and Model 3 (w/ muons and w/ convection, dotted) found solving the Boltzmann equations (blue), adopting the maximum entropy method (purple), and the Minerbo closure (pink), as well as the linear (red), quadratic (orange), cubic 1 (light green), and cubic 2 (dark green) polynomial functions. Each panel represents a different post-bounce time snapshot, with the $y$-axis being chosen as arbitrary. 
  The gray vertical lines mark the energy-integrated decoupling radii of electron neutrinos, while the brown lines represent the shock radii, with solid, dashed, and dotted line styles corresponding to Model~1, Model~2, and Model~3, respectively. None of the approximate methods adopted to look for ELN crossings exactly match the ELN crossings found by solving the Boltzmann equations; see the main text for details. Note that the dots appearing for the maximum entropy method for $t_{\rm pb} = 0.05$, $0.1$ and $3$~s are due to the fact that ELN crossings are found in some radial bins, but not in the neighboring bins.
  }
  \label{fig:crossings-all}
\end{figure*}

\subsection{Results: Comparison among different methods}
We now compare the performance of the methods introduced in Sec.~\ref{Sec:methods} to identify ELN crossings in Models 1, 2, and 3 at the six selected post-bounce time snapshots. 
In order to identify ELN crossings, when solving the Boltzmann equations, we apply the selection criteria listed in Sec.~\ref{Subsec:Boltzmann}. 
For consistency, the same selection criteria are applied to the Minerbo and the maximum entropy schemes, although angular resolution is not a concern in these cases.

Figure~\ref{fig:crossings-all} summarizes our findings on the locations of the ELN crossings. The different line colors represent the methods introduced in Sec.~\ref{Sec:methods}, while the line styles distinguish among SN models, and each panel corresponds to a different post-bounce time snapshot. 
The energy-integrated decoupling radius of electron neutrinos (defined as the radius where the flux factor $\tilde{f} \simeq {1}/{3}$, see Eq.~\ref{Eq:flux_factor}) as well as the shock radius are marked by vertical lines to guide the eye. 
We stress that we select only forward crossings for the angular distributions computed from the Boltzmann equations (blue) and the maximum entropy method (purple).

The angular distributions computed by solving the Boltzmann equations (blue lines) are in agreement with the ones of Refs.~\cite{Shalgar:2024gjt,Shalgar:2025oht} for the same SN model with PNS convection and without muons (analogous of our Model 2--dashed lines--but for the Steiner, Fischer, and Hempel (SFHo) equation of state~\cite{2013ApJ...774...17S}). Note, however, that Refs.~\cite{Shalgar:2024gjt,Shalgar:2025oht} do not find crossings at $t_{\mathrm{pb}}<0.5$ s; these differences arise from slight variations in the input physics and equation of state. For all post-bounce times, we find ELN crossings above the neutrinosphere. Nevertheless, the presence of such crossings depends on the SN model and post-bounce time. In fact, Ref.~\cite{Cornelius:2025nvd} shows that PNS convection tends to weaken crossings, favoring smaller differences between the decoupling radii of electron neutrinos and antineutrinos and making their angular distributions more similar to each other.
As a consequence, Models 2 and 3 (w/ convection) tend to display ELN crossings starting at radii larger than the ones found for Model 1 (w/o convection). At $t_{\mathrm{pb}}=3$~s, no ELN crossings are detected except for Model 3. This finding is a hint that flavor instabilities are less prominent in the late cooling phase.

The maximum entropy method (purple lines in Fig.~\ref{fig:crossings-all}) leads to ELN crossings in spatial regions roughly overlapping with the results from the Boltzmann equations, for $t_{\rm pb} \lesssim 1$~s.
Interestingly, the differences between the findings of the maximum entropy approach and the Boltzmann solution are more pronounced for Models 2 (dashed lines) and 3 (dotted lines) at $t_{\rm pb} = 0.1$ and $3$~s. 
Moreover, at those post-bounce times Models 2 and 3 alternate radii at which ELN crossings exist to radii where they do not (hence the purple line for Model 2 appears dotted instead of dashed). The same occurs at $t_{\rm pb} = 0.05$~s at large radii for all models.

The Minerbo method (pink lines in Fig.~\ref{fig:crossings-all}) performs significantly worse than the maximum entropy scheme. Only for Model 2 (Model 3) at $t_{\rm pb} = 1$~s ($t_{\rm pb} = 3$~s) most of the crossings found through the Boltzmann solution are also identified by means of the Minerbo method.
At all other post-bounce time snapshots, none of the crossings from the Boltzmann solution are found by the Minerbo approach. 

Figure~\ref{fig:crossings-all} shows that the polynomial method (red, orange, light green and dark green lines) performs worse than the maximum entropy method, when compared to the Boltzmann solution. The linear (red) and quadratic polynomials (orange) detect crossings for $t_{\rm{pb}} \gtrsim 0.5$~s. For example, for $t_{\rm pb} = 0.5$~s, the ELN distribution drops sharply around the crossing (cf.~Fig.~\ref{fig:ang_dists}); this rapid decline of $G(r,\mu)$ around the crossing makes it easier to find a function $\mathcal{F}(\mu)$ that leads to $I_{\mathcal{F}}(r)I_0(r)<0$. However, the linear and quadratic polynomials fail to identify crossings for $t_{\rm pb}\lesssim 0.1$~s. The cubic polynomial 1 (light green), which was expected to perform better since it uses the first four moments, mainly detects crossings before decoupling. 
It is also interesting to note that the Cubic 1 and Cubic 2 functions find crossings in alternating spatial regions. This happens because Cubic 1 tends to highlight forward crossings, while Cubic 2 focuses on backward ones. As a result, they each detect crossings only in certain radial ranges, contrary to the Boltzmann results, which find crossings more consistently. This highlights how the performance of the polynomial method can be largely hampered by the wrong choice of polynomial function (which is difficult to gauge a priori, since when searching for ELN crossings in moment-based simulations, the shape of the angular distributions is not known).

\subsection{Discussion}

Figure~\ref{fig:crossings-all} highlights that while forward ELN crossings are found for all post-bounce times when solving the Boltzmann equations, moment-based methods do not allow reliably diagnosing the presence of ELN crossings. This is due to the fact that the higher angular moments of the neutrino distribution carry important information on the structure of the angular distributions that is otherwise cut when approximating the neutrino field with the first few moments. These general conclusions are in agreement with those of Ref.~\cite{Johns:2021taz}. 

A large body of work employs the maximum entropy approach to reconstruct neutrino angular distributions. In agreement with our conclusions, Refs.~\cite{Abbar:2023zkm, Johns:2021taz} argued that while the maximum entropy closure captures some key features of the angular distributions more accurately than the bulb model or the Minerbo closure, it is not the optimal choice to diagnose the existence of ELN crossings. Machine learning models trained on angular distributions computed without approximations outperform with respect to distributions reconstructed relying on a few moments~\cite{Abbar:2023kta,Abbar:2023zkm,Abbar:2024chh}. 
Reference~\cite{Richers:2022dqa} focused on neutron-star merger remnant simulations and concluded that the maximum entropy method identifies ELN crossings with comparable performance to the one of Monte Carlo simulations. However, the fact that Ref.~\cite{Richers:2022dqa} does not distinguish between ELN crossings in the forward and backward directions might have led to the misidentification of some crossings because of the numerical fluctuations discussed in Sec.~\ref{Subsec:Boltzmann}. 
The limitations of the Minerbo approach to look for ELN crossings are also highlighted in Refs.~\cite{Johns:2021taz,Nagakura:2021txn}, where it is argued that closure schemes are unreliable to describe forward-peaked distributions.

The polynomial method was expected to detect only a fraction of all ELN crossings, since the method is limited to a maximum moment rank of $N=3$. Even so, our results show performance of the polynomial method which is worse than the $40\%$ success rate reported in Ref.~\cite{Abbar:2020fcl} (which attempted to detect through this method the forward crossings computed by solving the Boltzmann equations in Ref.~\cite{Abbar:2018shq}).
In agreement with our findings of the linear, quadratic and cubic 1 polynomials, Ref.~\cite{Richers:2022dqa} found that the chosen polynomials could not detect ELN crossings across the entire spatial range otherwise found through other methods. Thus, the author concluded that the maximum entropy method outperforms the polynomial method for the chosen weighting functions.

Reference~\cite{Abbar:2020qpi} tweaked the weighting functions to search for backward crossings above the shock and forward crossings below the shock. 
Following Ref.~\cite{Abbar:2020qpi}, we have designed cubic polynomial 2 (dark green) to capture the crossings not identified with the other polynomial functions (making it sensitive to backward crossings); in fact, this polynomial function has identified crossings in both the pre- and post-shock regions.
Similarly, we have employed cubic polynomial 1 to look for forward crossings, detecting crossings in the PNS region for all post-bounce times. However, these crossings are not found by the Boltzmann solution, but most of them appear for cubic polynomial 2. It is thus questionable whether the ELN crossings are forward. In the same way, it is unclear whether all the crossings identified by cubic polynomial 2 are backward since the Boltzmann solution finds forward crossings for parts of the radial range.
We conclude that it is doubtful whether tweaking the weighting functions can help in diagnosing the type of crossing; in Ref.~\cite{Abbar:2020qpi} they detect some crossings in the post-shock region that they claim could be both forward and backward, even though the polynomial used was tweaked to detect forward crossings in that region.
The misidentification of ELN crossings could lead to wrong conclusions about how significant flavor conversion would be, since backward crossings are unlikely to lead to flavor conversions due to the small backward neutrino flux.
The polynomial method may capture a number of ELN crossings in the backward direction, leading to an overestimation of the SN regions affected by flavor conversion, if used as a diagnostic method.

Another critical issue with the polynomial method is its inability to quantify the strength of ELN crossings. In fact, for what concerns the physical implications, not all crossings are equally significant; e.g.~a shallow backward crossing in the pre-shock region or in the outer post-shock layer may have a negligible impact on neutrino flavor evolution.
As noted in Ref.~\cite{Johns:2021taz}, the polynomial method likely does not lead to wrong identification of crossings. However, an inappropriate choice of the polynomial function may misleadingly suggest conditions for flavor conversion to occur over a large portion of the radial domain. Cubic polynomial 2 was used to test whether a weighting function could detect crossings at all radii. While it nearly achieves this goal by identifying crossings inside the PNS and after decoupling for all models and time steps except at $t_{\rm pb}=0.5$ s and $0.75$ s, many of these crossings do not appear in searches employing other approaches. This suggests that most of these crossings must be insignificant.

\section{Conclusions}
\label{Sec:conclusions}
Because of the computational costs, most existing hydrodynamical simulations of core-collapse SNe do not provide flavor-dependent angular distributions of neutrinos and antineutrinos, with neutrino transport being often approximated relying on the first few moments. However, the shape of the neutrino angular distributions, and in particular of the ELN distribution, is crucial to model flavor conversion. 

In this paper, we compare four different approaches adopted in the literature to look for the existence of ELN crossings in state-of-the-art SN simulations: the solution of the Boltzmann equations, the maximum entropy method, the Minerbo closure, and the polynomial method. We applied these methods to three spherically symmetric SN models for a progenitor of $18.6 M_\odot$: Model 1 (without muons and without convection), Model 2 (without muons and with convection), and Model 3 (with muons and with convection), relying on six post-bounce times spanning both the accretion and cooling phases of SN evolution.
The solution of the Boltzmann equations demonstrates that ELN crossings are present in all models and at multiple time steps. However, if the angular distributions from the Boltzmann solutions are not used, but instead one reconstructs the angular distributions through moment-based schemes, the specific locations and detectability of these crossings strongly depend on the method used.

We consider the angular distributions computed by solving the Boltzmann equations as our benchmark since their computation does not involve any approximation, except for limitations due to numerical discretization. The maximum entropy method successfully identifies crossings in certain regions of the SN core, but it fails in identifying crossings in parts of the explored radial range for some of the investigated time snapshots or identifies crossings where the Boltzmann solution does not find any.
The polynomial method performs the worst overall, with linear and quadratic polynomials only proving roughly successful between $t_{\mathrm{pb}}=0.5$~s and $1$~s
and the cubic polynomials finding crossings either exclusively inside the PNS or close to it (cubic polynomial 1) or simply finding crossings in most of the radial range (cubic polynomial 2). Similarly, the Minerbo closure only finds crossings for $t_{\mathrm{pb}}=1$~s and $3$~s and only for a subset of our models.

The trend emerging from this comparison highlights the limitations inherent in the reconstruction of the angular distributions relying only on low-order moments. Similar conclusions have also been reported in Refs.~\cite{Johns:2021taz,Nagakura:2025hss}.
Our results have important implications for investigating neutrino flavor conversion in SNe as well as related searches for ELN crossings. Since neutrino quantum kinetics can play an important role for the explosion mechanism and the nucleosynthetic outcome of SNe~\cite{Ehring:2023abs,Ehring:2023lcd,Nagakura:2023mhr,Wang:2025nii}, the inference of the regions of flavor conversion relying on the first few moments could lead to a wrong assessment of the impact of neutrino quantum kinetic effects on SN physics. 
These conclusions also apply to neutron star merger remnants, where it has been found that neutrino flavor conversion could affect the disk cooling as well as boost the fraction of lanthanides relying on neutrino transport methods employing different degrees of approximation~\cite{Wu:2017drk,Just:2022flt,George:2020veu,Lund:2025jjo,Qiu:2025kgy}.

\section{Acknowledgments}
We thank Sajad Abbar for useful discussions and Shashank Shalgar for contributing to the development of the Boltzmann solver adopted in Ref.~\cite{Cornelius:2025nvd} whose output is employed in this paper.
At the Niels Bohr Institute, this project has received support from the Villum Foundation (Project No.~13164) and the European Union (ERC, ANET, Project No.~101087058). 
In Garching, the work was supported by the German Research Foundation (DFG) through the Collaborative Research Centre ``Neutrinos and Dark Matter in Astro- and Particle Physics (NDM),'' Grant No.~SFB-1258-283604770, and under Germany's Excellence Strategy through the Cluster of Excellence ORIGINS EXC-2094-390783311. 
Views and opinions expressed are those of the authors only and do not necessarily reflect those of the European Union or the European Research Council. Neither the European Union nor the granting authority can be held responsible for them. The Tycho supercomputer hosted at the SCIENCE HPC Center at the University of Copenhagen was used to support the numerical simulations presented in this work. Computing resources are also acknowledged from the Max Planck Computing and Data Facility (MPCDF) on the HPC systems Cobra, Draco, and Raven.

\appendix
\section{Inputs adopted in the polynomial method}
\label{Sec:appendix}
\begin{figure}
  \centering
  \includegraphics[width=0.82\linewidth]{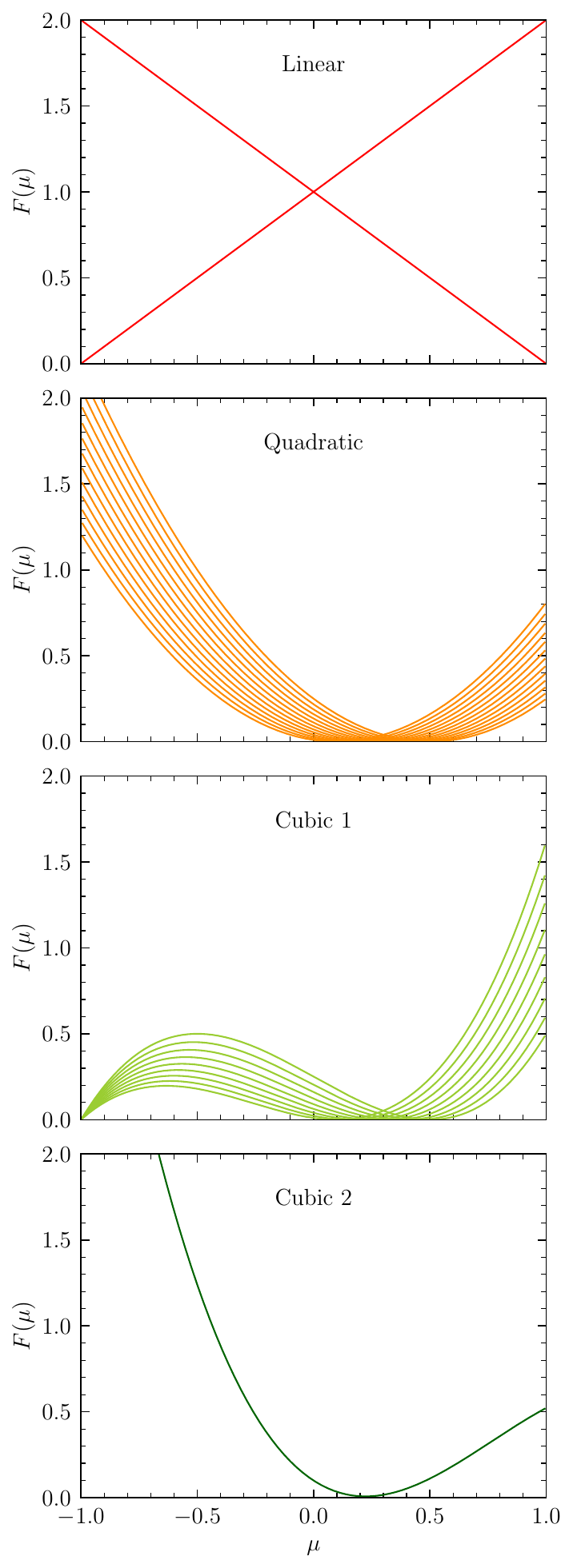}
  \caption{Functions adopted in the polynomial method.}
  \label{fig:polynomials}
\end{figure}
Figure~\ref{fig:polynomials} illustrates the linear, quadratic, and cubic polynomials employed in this work to search for ELN crossings by relying on the polynomial method. We mimic the shapes of the quadratic (orange) and cubic polynomials (cubic 1, light green) to those considered suitable to detect forward crossings in Ref.~\cite{Abbar:2020qpi} in the post-shock region where such crossings are expected. We express the quadratic functions as $\mathcal{F}(\mu)=(\mu-d)^2$ and the cubic 1 functions as $\mathcal{F}(\mu)=(\mu-d)^2(\mu+1)$ where $d$ is the local minimum.
To maximize the detection of ELN crossings, the local minimum of each polynomial should be located near an expected peak of the ELN distribution (for $\mu>0$).
This approach ensures a higher probability of capturing as many forward ELN crossings as possible. The crossings reported in Fig.~\ref{fig:crossings-all} are the combined results of the polynomials of each type.
We also include a second type of cubic polynomial (cubic 2, dark green) to cross-check whether there could exist a polynomial function capturing crossings otherwise not detected through the other polynomial functions. 

\bibliography{references}

\end{document}